\documentstyle[psfig,prb,aps,multicol]{revtex}

\draft
\begin{document}

\title{Mixed spin ladders with exotic ground states}

\author{A. K. Kolezhuk\cite{perm}}
\address{Institut f\"{u}r Theoretische Physik,
Universit\"{a}t Hannover, Appelstr. 2, 30167 Hannover, Germany\\
and Institute of Magnetism, National Academy of
Sciences and Ministry of Education of Ukraine, \\
36(b) Vernadskii avenue, 252142 Kiev, Ukraine}

\author{H.-J. Mikeska}
\address{Institut f\"{u}r Theoretische Physik,
Universit\"{a}t Hannover, Appelstr. 2, 30167 Hannover, Germany}

\date{November 25, 1997}
\maketitle

\begin{abstract}
We study the ``mixed spin'' isotropic ladder system having $S=1$ spins
on one leg and $S={1\over2}$ spins on the other, with general-type
exchange interactions between spins on neighboring
rungs. A set of model Hamiltonians with exact ground states in the
form of a certain matrix product wave function is obtained.  We show
that sufficiently strong frustration can lead to exotic singlet ground
states with infinite (exponential) degeneracy. We also list a couple
of rather simple models with nontrivial ground states, including a
model with only bilinear exchange.
\end{abstract}


\maketitle

\begin{multicols}{2}

\section{Introduction}
\label{sec:intro}

During the last few years, ``mixed'' one-dimensional (1d) quantum spin
systems composed of two or more different kinds of spin have drawn
certain interest. Several Bethe-ansatz solvable models with singlet
ground states have been found;
\cite{AndreiJohanneson84,DeVegaWoynarovich92,DeVega+94,
Schlottmann94,ZvyaginSchlottmann95,Fuji+96,DorfelMeissner97}
experimentally relevant models (see, e.g., Ref. \onlinecite{Kahn+92})
of ferrimagnetic quantum chains were studied
\cite{Pati+97,BreMikYam97,KMY97} via various numerical and analytical
approaches; recently, mixed spin chains
\cite{FukuiKawakami97,Tonegawa+97} and ladders
\cite{FukuiKawakami-preprint} with singlet ground states were analyzed
both numerically and by means of the nonlinear sigma model technique,
and other interesting mixed-spin models with rich phase diagram were
proposed and investigated. \cite{NiggUimZitt97}

On the other hand, there has been a considerable progress in studying
1d spin systems by means of the so-called matrix product (MP) states
technique.  \cite{Fannes+89,Klumper+91,Fannes+92,Klumper+92-93} MP
states have proved to be a particularly useful tool for constructing
new models with exact ground states: ground states of the MP type were
found for a large family of spin-1 and spin-$3\over2$
chains\cite{Klumper+91,Fannes+92,Klumper+92-93,Lange+94,NiggZitt96}
and spin-$1\over2$ ladders. \cite{GSu96,KM97} The simplest example of
the MP state is the spin-1 valence bond state (VBS) which is the
ground state of the AKLT model \cite{AKLT} and is widely accepted as a
convenient image of the Haldane-phase state.  MP states were also
successfully used for the variational study of the ground state and
elementary excitations of spin chains
\cite{SchadZitt95,TotsukaSuzuki95,NeuMik96} and ladders.
\cite{Brehmer+96,Brehmer+97} However, the states which can be accessed
via the MP approach have finite, typically rather short, correlation
length. Therefore they are usually gapped (except for the case of
ferromagnetic-type situation with spontaneously broken symmetry
\cite{KMY97,Oshikawa92} when the system has long-range order and is
gapless due to the Goldstone modes) and cannot be applied for a
description of the system behavior at any critical point.

\begin{figure}
\mbox{\psfig{figure=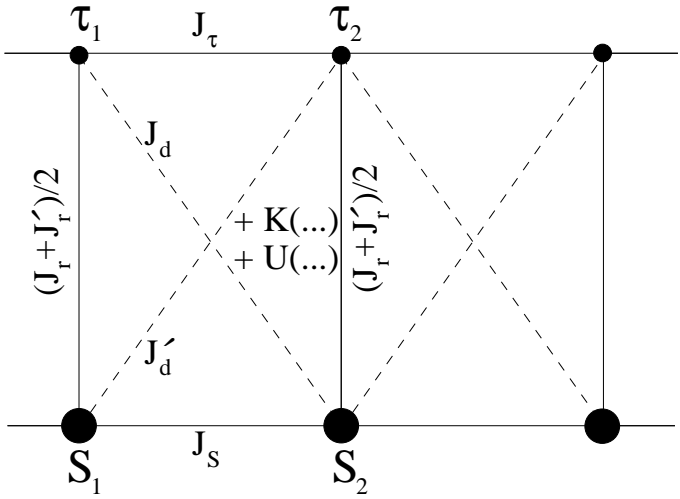,width=90mm,angle=-90}}
\vskip 5mm
\caption{Mixed spin ladder with the $S={1\over2}$ upper leg and
$S=1$ lower leg as described by the model (\protect\ref{ham}). }
\label{fig:mixedlad}
\end{figure}

In this paper, we study the isotropic mixed-spin system which may be
viewed as a ladder composed of $S=1$ and $S={1\over2}$ legs, with
general bilinear, biquadratic, and six-spin interactions between
neighboring {\em rungs,\/} as shown in Fig. \ref{fig:mixedlad}.  The
model is described by the Hamiltonian $\widehat H=\sum_n \widehat
h_{n,n+1}$, where $\widehat h$ is chosen in the following form:
\begin{eqnarray}
\widehat h_{12}&=&
J_S(\mbox{\boldmath$S$\unboldmath}_1\mbox{\boldmath$S$\unboldmath}_2)
+J_\tau(\mbox{\boldmath$\tau$\unboldmath}_1\mbox{\boldmath$\tau$\unboldmath}_2)
\nonumber\\
&+&{1\over2}J_r
(\mbox{\boldmath$S$\unboldmath}_1\mbox{\boldmath$\tau$\unboldmath}_1)
+{1\over2}J_r'
(\mbox{\boldmath$S$\unboldmath}_2\mbox{\boldmath$\tau$\unboldmath}_2)
+J_d(\mbox{\boldmath$\tau$\unboldmath}_1\mbox{\boldmath$S$\unboldmath}_2)
+J_{d}'(\mbox{\boldmath$S$\unboldmath}_1\mbox{\boldmath$\tau$\unboldmath}_2)
\nonumber\\
&+&K_S
(\mbox{\boldmath$S$\unboldmath}_1\mbox{\boldmath$S$\unboldmath}_2)^2
+K_{S\tau}
(\mbox{\boldmath$S$\unboldmath}_1\mbox{\boldmath$S$\unboldmath}_2)
(\mbox{\boldmath$\tau$\unboldmath}_1\mbox{\boldmath$\tau$\unboldmath}_2)
\nonumber\\
&+&K_{rr}
(\mbox{\boldmath$S$\unboldmath}_1\mbox{\boldmath$\tau$\unboldmath}_1)
(\mbox{\boldmath$S$\unboldmath}_2\mbox{\boldmath$\tau$\unboldmath}_2)
+K_{dd}
(\mbox{\boldmath$\tau$\unboldmath}_1\mbox{\boldmath$S$\unboldmath}_2)
(\mbox{\boldmath$S$\unboldmath}_1\mbox{\boldmath$\tau$\unboldmath}_2)
\nonumber\\
&+&K_{rd}
(\mbox{\boldmath$S$\unboldmath}_1\mbox{\boldmath$\tau$\unboldmath}_1)
(\mbox{\boldmath$S$\unboldmath}_1\mbox{\boldmath$\tau$\unboldmath}_2)
+K_{rd}'(\mbox{\boldmath$S$\unboldmath}_2\mbox{\boldmath$\tau$\unboldmath}_2)
(\mbox{\boldmath$\tau$\unboldmath}_1\mbox{\boldmath$S$\unboldmath}_2)
\nonumber\\
&+&(\mbox{\boldmath$S$\unboldmath}_1\mbox{\boldmath$S$\unboldmath}_2) 
\big[
K_{1}(\mbox{\boldmath$S$\unboldmath}_1\mbox{\boldmath$\tau$\unboldmath}_1)
+K_{1}'(\mbox{\boldmath$S$\unboldmath}_2\mbox{\boldmath$\tau$\unboldmath}_2)
\big]
\nonumber\\
&+&(\mbox{\boldmath$S$\unboldmath}_1\mbox{\boldmath$S$\unboldmath}_2)
\big[
K_{2}(\mbox{\boldmath$\tau$\unboldmath}_1\mbox{\boldmath$S$\unboldmath}_2)
+K_{2}'(\mbox{\boldmath$S$\unboldmath}_1\mbox{\boldmath$\tau$\unboldmath}_2)
\big]
\nonumber\\
&+&U_1
(\mbox{\boldmath$S$\unboldmath}_1\mbox{\boldmath$S$\unboldmath}_2)^2
(\mbox{\boldmath$\tau$\unboldmath}_1\mbox{\boldmath$\tau$\unboldmath}_2)
+U_2
(\mbox{\boldmath$S$\unboldmath}_1\mbox{\boldmath$\tau$\unboldmath}_1)
(\mbox{\boldmath$S$\unboldmath}_1\mbox{\boldmath$S$\unboldmath}_2)
(\mbox{\boldmath$S$\unboldmath}_2\mbox{\boldmath$\tau$\unboldmath}_2)
\nonumber\\
&+&U_3
(\mbox{\boldmath$\tau$\unboldmath}_1\mbox{\boldmath$S$\unboldmath}_2)
(\mbox{\boldmath$S$\unboldmath}_1\mbox{\boldmath$S$\unboldmath}_2)
(\mbox{\boldmath$S$\unboldmath}_1\mbox{\boldmath$\tau$\unboldmath}_2)
-E_{0}\cdot \widehat 1\,.
\label{ham}
\end{eqnarray}
Here $\mbox{\boldmath$ S$\unboldmath}$ and
$\mbox{\boldmath$\tau$\unboldmath}$ denote spin-1 and spin-$1\over2$
operators, respectively, and symmetric ordering of spin-1 operators is
assumed wherever it is necessary. 

We construct different {\em
singlet\/} MP wave functions interpolating between a few simple VBS
states, and use the technique of ``optimal ground states''
\cite{NiggZitt96} to find a family of Hamiltonians for which those
wave functions are exact ground states.  Among the members of this
family, the following interesting representatives are found:

(i) {\em ``AKLT-like'' models\/} which differ from the AKLT model by a few
additional terms;

(ii) {\em biquadratically coupled chains\/} which do not
contain bilinear exchange interactions between the $S=1$ and $S={1\over2}$
legs;

(iii) a model with {\em  purely bilinear interactions};

(iv) {\em multicritical models\/} with infinitely degenerate singlet
ground states (for finite systems the degeneracy is exponentially
large). 

By the construction of our MP ansatz, all the ground states are
dimerized, and since we choose the initial Hamiltonian to be
translationally invariant, they are {\em spontaneously\/} dimerized.
This is in line with the recent field-theoretical argument
\cite{NersesyanTsvelik97} that sufficiently strong biquadratic
interaction can cause spontaneous dimerization in $S={1\over2}$
ladder, implying also a similar two-particle structure of the spectrum
(``absence of magnons''). [The elementary excitation of a
spontaneously dimerized system is a pair of domain walls in the dimer
order.]

The paper is organized as follows: Sect. \ref{sec:mp} explains the
principles of constructing the matrix product ansatz and the procedure
of finding the set of exact solutions, Sect. \ref{sec:results}
contains the most important results describing representative models
as listed above, and Sect. \ref{sec:summary} gives a brief
summary. For convenience, technical details and general
solutions are listed in the Appendices, so that those readers who are not
interested in the details may go directly to Sect. \ref{sec:results}.

\section{The matrix product ansatz and optimal ground states
construction}
\label{sec:mp}

We construct the MP ansatz for the model (\ref{ham}) from rectangular
matrices $g^{L}$ and $g^{R}$ as follows:
\begin{equation}
\label{mp}
|\Psi_{0}\rangle = \mbox{Tr}(g^{L}_{1}g^{R}_{2}g^{L}_{3}g^{R}_{4}
\cdots g^{L}_{2N-1}g^{R}_{2N})\,,
\end{equation}
where the matrix $g^{L,R}_{i}$ contains spin states of the $i$-th rung
only, and the total number of rungs is $2N$. We demand that
$|\Psi_{0}\rangle$ is a global singlet, then
according to the approach proposed in Ref. \onlinecite{KMY97} this dictates
the following structure of the elementary matrices $g^{L,R}$:
\begin{eqnarray}
\label{ansatz}
g^{L,R}&=&\sum_{k,\lambda={1\over2},{3\over2}}\sum_{q,\mu} c^{(k,\lambda)} 
\langle 00|kq,\lambda\mu\rangle T^{kq}_{L,R} |\psi_{\lambda\mu}\rangle
\\
&=&a\, {1\over\sqrt{2}} \left(T^{{1\over2},{1\over2}}_{L,R}
|\Downarrow\rangle -T^{{1\over2},-{1\over2}}_{L,R} |\Uparrow\rangle\right)
\nonumber\\
&+&b\, {1\over2} \Big( T^{{3\over2},{3\over2}}_{L,R}
|\bar{3}\rangle  -
T^{{3\over2},-{3\over2}}_{L,R}
|3\rangle -
T^{{3\over2},{1\over2}}_{L,R}
|\bar{1}\rangle
+T^{{3\over2},-{1\over2}}_{L,R}
|1\rangle
\Big).\nonumber
\end{eqnarray}
Here $T_{L,R}^{kq}$ are in general arbitrary matrix representations of
irreducible tensor operators $\widehat{T}^{kq}$ transforming under
rotations $\widehat{R}$ according to the ${\cal D}^{k}$ representation
of the rotation group:
\[
\widehat{R}\, \widehat{T}^{kq} = \sum_{q'}{\cal
D}^{k}_{q'q}(\widehat{R})  \widehat{T}^{kq'}\,,
\]
and $|\psi_{\lambda\mu}\rangle$ are the rung wave functions with total spin
$\lambda$ and its $z$-projection $\mu$.  The quantities $a\equiv
c^{({1\over2},{1\over2})}$ and $b\equiv c^{({3\over2},{3\over2})}$ are
free parameters.  We use the compact notation $|\Uparrow\rangle$,
$|\Downarrow\rangle$ for the rung states having $\lambda={1\over2}$, and
$|1\rangle$, $|\bar{1}\rangle$, $|3\rangle$, $|\bar{3}\rangle$ for the
states with spin $\lambda={3\over2}$:
\begin{eqnarray}
\label{rung-states}
&&|\Uparrow\rangle  = \sqrt{2\over3} |+\downarrow\rangle
-{1\over\sqrt{3}} |0\uparrow\rangle,\;\;
   |\Downarrow\rangle  =
{1\over\sqrt{3}} |0\downarrow\rangle
-\sqrt{2\over3} |-\uparrow\rangle
,\nonumber\\
&&|3\rangle =
|+\uparrow\rangle,\quad
|\bar{3}\rangle = |-\downarrow\rangle,\\
&&|1\rangle = \sqrt{2\over3} |0\uparrow\rangle
+{1\over\sqrt{3}} |+\downarrow\rangle,\quad
  |\bar{1}\rangle  =\sqrt{2\over3} |0\downarrow\rangle
+{1\over\sqrt{3}} |-\uparrow\rangle\,;\nonumber
\end{eqnarray}
here single arrows $|\uparrow\rangle$, $|\downarrow\rangle$
indicate the spin-$1\over2$ states, and $|+\rangle$, $|0\rangle$,
$|-\rangle$ denote spin-$1$ states.

The matrix elements of any irreducible tensor operator (for
definiteness, let us choose $\widehat{T}^{kq}_{L}$) in a fixed basis,
according to the Wigner-Eckart theorem, are given by
\begin{eqnarray}
\label{TL2x3}
T^{kq}_{L}(M',M)&\equiv& \langle JM| \widehat{T}^{kq}_{L} |J'M'\rangle
\nonumber\\
&=&\widetilde{T}^{k}_{J,J'}\, \langle JM | kq,J'M'\rangle\,, 
\end{eqnarray}
where the reduced matrix element $\widetilde{T}^{k}_{J,J'}$ does not
depend on $M$, $M'$, $q$ and thus can be absorbed into the free
parameters $c^{(k\lambda)}$ in (\ref{ansatz}).  We need $k$ to be
half-integer, therefore necessarily $J\not=J'$ and one arrives at a
$(2J+1)\times(2J'+1)$ non-square matrix. Further, we fix the choice of
$\widehat{T}^{kq}_{R}$ defining it as
\begin{equation} 
\label{TR} 
(\widehat{T}^{kq}_{L})^{\dag}=(-)^{k-q}\, \widehat{T}^{k,-q}_{R}\,.
\end{equation}
For our problem one can choose, e.g., $J=1$ and $J'={1\over2}$, then
we have $3\times2$ and $2\times3$ matrices for $T_{L}$ and $T_{R}$,
respectively, so that the dimension of matrix space coincides with the
total number of states of one rung. 

More complicated basis for $T^{kq}_{L}$ can be chosen, e.g., one may
``decompose'' $J=1$ into two $J_{1}=J_{2}={1\over2}$ and define
\begin{eqnarray} 
\label{TL2x4} 
&&T_{L}^{kq,J_{12}}(M',M_{1} M_{2}) \equiv \langle J_{1}M_{1} J_{2}M_{2}
|T^{kq}_{L}|J_{12}JM\rangle \nonumber\\
&& \qquad\qquad = \widetilde{T}^{k}_{J,J_{1},J_{2},J_{12}} \,
\langle J_{12} JM|kq,J_{1}M_{1},J_{2}M_{2}\rangle\,,
\end{eqnarray}
where $J_{12}$ denotes the eigenvalues of the operator 
$(\widehat{{\mathbf J}}_{1} +\widehat{{\mathbf J}}_{2})^{2}$.  
Then, combining $M_{1}$ and
$M_{2}$ into one ``composite'' index, one gets $4\times2$ matrices for
$T^{kq}_{L}$ and $4\times2$ respectively for $T^{kq}_{R}$.  Another
difference with the previous case is that now for $k={1\over2}$ there
are two possibilities: $J_{12}=1$ or $0$, and the set of matrices
$T^{{1\over2}q}$ acquires an additional free parameter $w$:
\[ 
T^{{1\over2}q}_{L,R}=T^{{1\over2}q,1}_{L,R}+ wT^{{1\over2}q,0}_{L,R}\,.
\]

\begin{figure}
\mbox{\psfig{figure=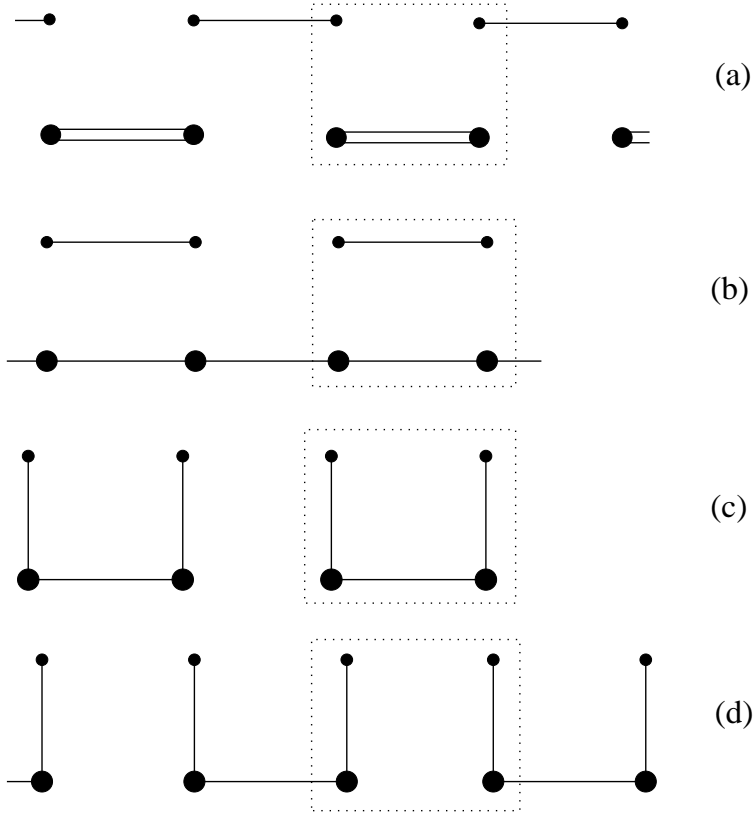,width=100mm,angle=-90}}
\vskip 10mm
\caption{A couple of VBS states which can be obtained as limiting
cases of the $2\times4$ and $2\times3$ MP ans\"atze
(\protect\ref{ansatz}). Dotted box corresponds to the matrix product
$g^{L}g^{R}$.}
\label{fig:VBS} 
\end{figure}

For the sake of brevity, we will further refer to those two
construction as ``$2\times3$'' and ``$2\times4$'' MP ans\"{a}tze; the
explicit form of the matrices we used can be found in the Appendix
\ref{app:tech} [see Eqs. (\ref{tjm2x3}) and (\ref{tjm2x4})].  One may
think of the above construction as of interpolating between several
VBS states shown in Fig. \ref{fig:VBS}. It is easy to check, for
example, that the $2\times4$ ansatz at ($b/a=\sqrt{2}$, $w=0$) and
$2\times3$ ansatz at $b/a=\sqrt{2}$ correspond to the same state with
completely dimerized legs  shown in Fig. \ref{fig:VBS}a, another choice of
parameters ($a/b=-2\sqrt{2}$, $w=\pm\sqrt{3}$) leads to the state with
$S={1\over2}$ leg dimerized and $S=1$ leg in the AKLT-type VBS state
(Fig. \ref{fig:VBS}b), and, finally, the combinations ($w=\pm 1/\sqrt{3}$,
$b=0$) and ($b=0$, $a\to0$, $w \propto 1/a$) give two ``U-shape''
dimerized states shown in Fig. \ref{fig:VBS}c,d respectively.

Our procedure of constructing the exact ground states follows the
ideas presented in
Refs. \onlinecite{Klumper+91,Klumper+92-93,NiggZitt96}: we require the
MP wave function (\ref{mp}) to be a zero-energy groundstate of the
local Hamiltonian $\widehat{h}_{i,i+1}$, which ensures that it is a
ground state of the global Hamiltonian $\widehat{H}$ (an {\em optimal}
ground state, in the terminology of Ref. \onlinecite{NiggZitt96}). This
yields the following conditions:

(i) $\widehat{h}_{i,i+1}$ annihilates all states being  matrix
elements of the two products $g^{L}_{i}g^{R}_{i+1}$ and
$g^{R}_{i}g^{L}_{i+1}$:
\begin{equation} 
\label{cond1} 
\widehat{h}_{12} (g^{L}_{1}g^{R}_{2})=0\,,\quad 
\widehat{h}_{12} (g^{R}_{1}g^{L}_{2})=0\,;
\end{equation}

(ii) all other eigenstates of $\widehat{h}_{12}$ have the energy
$\varepsilon\geq 0$. Then $|\Psi_{0}\rangle$ is the zero-energy ground state
of $\widehat{H}$; if one drops the constant term $-E_{0}\cdot
\widehat{1}$ in (\ref{ham}), the remaining Hamiltonian has the energy
density $E_{0}$ per rung.

For further treatment it is convenient to write the Hamiltonian
$\widehat{h}$ in terms of projectors on the states
$|\Psi_{JM}^{(k)}\rangle$ of the two-rung plaquette $(i,i+1)$ with
fixed angular momentum. The complete set of the plaquette states
[see Eqs. (\ref{plaquette-states})]
contains one multiplet with $J=3$, three quintuplets, four triplets and
two singlets, and thus one
obtains:
\begin{eqnarray} 
\label{ham-proj}
&&\widehat{h}= \!\!\!\sum_{k,l=1,2} \!\! \lambda_{0}^{(k,l)}
|\Psi_{00}^{(k)}\rangle \langle \Psi_{00}^{(l)}| + \!\!\!
\sum_{k,l=1..4} \!\! \lambda_{1}^{(k,l)}\sum_{M}
|\Psi_{1M}^{(k)}\rangle \langle \Psi_{1M}^{(l)}|\nonumber\\
&&\;+ \!\!\! \sum_{k,l=1..3} \!\!\!\lambda_{2}^{(k,l)}\sum_{M}
|\Psi_{2M}^{(k)}\rangle \langle \Psi_{2M}^{(l)}| +
\lambda_{3}\sum_{M}
|\Psi_{3M}\rangle \langle \Psi_{3M}|\,.
\end{eqnarray}
Here obviously $\lambda_{J}^{(k,l)}=(\lambda_{J}^{(l,k)})^{*}$ because
of the hermitian property of $\widehat{h}$. 
The complete set of the
plaquette states can be divided into two subsets: {\em local ground
states\/} $|\Psi_{JM}^{g,(k)}\rangle$, $k=1,\ldots n_{J}^{(g)}$ which are
contained in the matrix products $g^{L}g^{R}$ and $g^{R}g^{L}$, and
{\em local eigenstates\/} $|\Psi_{JM}^{e,(k)}\rangle$, $k=1,\ldots
n_{J}^{(e)}$ which do not enter there. 

The conditions (i) mean that the local Hamiltonian $\widehat{h}$
should project only onto the states $|\Psi_{JM}^{e,(k)}\rangle$, and
the multiplets $|\Psi_{JM}^{g,(k)}\rangle$ have to be absent in Eq.\
(\ref{ham-proj}). This results in the following system of equations:
\begin{eqnarray} 
\label{zeroeqs} 
&\lambda_{J}^{(g,(k); g,(l))}=0,&
\;\;k=1\ldots n_{J}^{(g)},\;\; l=k\ldots n_{J}^{(g)}\,,\nonumber\\ 
&\lambda_{J}^{(e,(k'); g,(k))}=0,&
\;\;k=1\ldots n_{J}^{(g)},\;\; k'=1\ldots n_{J}^{(e)}\,,
\end{eqnarray}
which is essentially a system of linear equations in the
Hamiltonian coupling constants $J_{..}$, $K_{..}$, $U_{..}$ [see Eq.\
(\ref{ham})]. 

The conditions (ii) require that all the eigenstates within the
subspace determined by the basis $|\Psi_{JM}^{e,(k)}\rangle$ have positive
energy, which yields the inequalities
\begin{equation} 
\label{lambdaeqs}
\widetilde{\lambda}_{J}^{(\alpha)} \geq 0\,,\quad \alpha=1\ldots n_{J}^{(e)}\,, 
\end{equation}
where $\widetilde{\lambda}_{J}^{(\alpha)}$ denotes the eigenvalues of
the matrix
\[
\left[
\begin{array}{lcl}
\lambda_{J}^{(1,1)} & \cdots &\lambda_{J}^{(1,n_{J}^{(e)})} \\
\vdots & \ddots & \vdots \\
\lambda_{J}^{(n_{J}^{(e)},1)} & \cdots &\lambda_{J}^{(n_{J}^{(e)},n_{J}^{(e)})}
\end{array}
\right]\,.
\]
If one or more of $\widetilde{\lambda}_{J}^{(\alpha)}$ is zero, this
may indicate an additional degeneracy of the ground state. 

Eqs.\ (\ref{ansatz}), (\ref{TL2x3}), (\ref{TR}), (\ref{TL2x4}) and
(\ref{cond1}), (\ref{zeroeqs}), (\ref{lambdaeqs}) will be the basis for the
further analysis.  Now we proceed to considering specific solutions of
those equations in various particular cases.

\section{Models with exact ground states}
\label{sec:results}

In this section we present a number of models with exact ground states
being the most simple representatives of different classes of
solutions mentioned in the Introduction. 

\subsection{${\mathbf 2\times3}$ MP ansatz}
\label{subsec:2x3}

\subsubsection{General case}

It is easy to verify that for the $2\times3$ MP ansatz the two matrix
products $g^{L}g^{R}$ and $g^{R}g^{L}$ contain generally only the
following multiplets:
\begin{eqnarray}
\label{lgs2x3} 
&& a^{2}\psi^{11}_{00}+(b^{2}/\sqrt{2})\psi^{33}_{00} \equiv
|\Psi_{00}^{(2)}\rangle\,, 
\nonumber\\
&& \psi^{11}_{1M}\equiv |\Psi_{1M}^{(2)}\rangle\,,\;\;
\psi^{33}_{1M}\equiv |\Psi_{1M}^{(3)}\rangle\,,
\nonumber\\
&&(1/\sqrt{2})(\psi^{31}_{1M}+\psi^{13}_{1M})\equiv
|\Psi_{1M}^{(2)}\rangle\,,
\nonumber\\
&&
b^{2}\psi^{33}_{2M} +\sqrt{2}ab (\psi^{31}_{2M}-\psi^{13}_{2M})\equiv
|\Psi_{2M}^{(3)}\rangle\,,
\end{eqnarray}
then the remaining multiplets can be chosen as
\begin{eqnarray} 
\label{les2x3} 
&&|\Psi_{00}^{(1)}\rangle=(b^{2}/\sqrt{2})
\psi_{00}^{11}-a^{2}\psi^{33}_{00}\,, \nonumber\\
&&|\Psi_{1M}^{(1)}\rangle =
{1\over\sqrt{2}}(\psi^{31}_{1M}-\psi^{13}_{1M}),
\;|\Psi_{2M}^{(1)}\rangle =
{1\over\sqrt{2}}(\psi^{31}_{2M}+\psi^{13}_{2M})\,,
\nonumber\\
&&|\Psi_{2M}^{(2)}\rangle=2ab\psi^{33}_{2M}-(b^{2}/\sqrt{2})
(\psi^{31}_{2M}-\psi^{13}_{2M})\,.
\end{eqnarray}
The conditions (\ref{zeroeqs},\ref{lambdaeqs}) now take the form 
\begin{eqnarray} 
\label{eqn2x3gen} 
&&\lambda_{0}^{(2,2)}=\lambda_{0}^{(1,2)}=\lambda_{0}^{(2,1)}=0\,,
\nonumber\\
&&\lambda_{1}^{(1,k)}
=\lambda_{1}^{(k,l)}=0,\quad 2\leq k\leq 4,\;\; k\leq l\leq4\,,
\nonumber\\
&& \lambda_{2}^{(1,3)}=\lambda_{2}^{(2,3)}=\lambda_{2}^{(3,3)}=0\,,\\
&&\lambda_{0}^{(1,1)}\geq 0,\; \lambda_{1}^{(1,1)}\geq 0,\;
\lambda_{3}\geq 0,\;\widetilde{\lambda}_{2}^{\alpha}\geq 0\,,\nonumber
\end{eqnarray}
where $\widetilde{\lambda}_{2}^{\alpha}$, $\alpha=1,2$ are the
eigenvalues of the matrix
\[
\left[
\begin{array}{lr}
\lambda_{2}^{(1,1)} & \lambda_{2}^{(1,2)} \\
\lambda_{2}^{(2,1)} & \lambda_{2}^{(2,2)} 
\end{array}
\right]\,.
\]
For the sake of  simplicity we now set
$\lambda_{2}^{(1,2)}=0$. In fact, one can show that this requirement
just fixes certain ``natural'' symmetries in (\ref{ham}), namely,
\begin{equation} 
\label{syms} 
J_{r}=J_{r}',\;\; J_{d}=J_{d}',\;\; K_{rd}=K_{rd}',\;\;
K_{1,2}=K_{1,2}'\,.
\end{equation}
Further, we require the six-spin couplings in (\ref{ham}) to be zero,
in order to make the model less cumbersome. Then one obtains the sytem
of nineteen equations [fifteen equations (\ref{eqn2x3gen}) and four
additional assumptions $\lambda_{2}^{(1,2)}=0$, $U_{1,2,3}=0$] for
twenty parameters of the Hamiltonian (\ref{ham}) and the free
parameter $u=a/b$ entering the $2\times3$ MP ansatz (\ref{ansatz}),
(\ref{TL2x3}). It turns out that one of those nineteen equations is
linearly dependent, and the general solution contains, beside $u$, two
additional free parameters $x$ and $y$. This solution in its general
form is presented in Appendix \ref{app:gen} [see Eqs.\
(\ref{sol2x3gen}), (\ref{ranges2x3})], and here we will just consider
its most interesting particular cases:

(a)  Setting $u=-{1\over4}\sqrt{2}$, $x={32\over27}y$ and fixing the
energy scale by choosing $y={1\over3}$, one gets the {\em ``AKLT-type''
model\/} of the form 
\begin{eqnarray} 
\label{AKLT-type1}
\widehat{H}&=&\sum_{n} 
\mbox{\boldmath$S$\unboldmath}_{n}\mbox{\boldmath$S$\unboldmath}_{n+1}
+{1\over3}\left(
\mbox{\boldmath$S$\unboldmath}_{n}\mbox{\boldmath$S$\unboldmath}_{n+1}
\right)^{2}\nonumber\\
&+&{1\over3}\mbox{\boldmath$S$\unboldmath}_{n} 
(2\mbox{\boldmath$\tau$\unboldmath}_{n}
+\mbox{\boldmath$\tau$\unboldmath}_{n-1}
+\mbox{\boldmath$\tau$\unboldmath}_{n+1})
\\
&+&{1\over3}(\mbox{\boldmath$S$\unboldmath}_{n}
\mbox{\boldmath$S$\unboldmath}_{n+1}) 
\big[
(\mbox{\boldmath$S$\unboldmath}_{n}
+\mbox{\boldmath$S$\unboldmath}_{n+1})
\cdot
(\mbox{\boldmath$\tau$\unboldmath}_{n}+
\mbox{\boldmath$\tau$\unboldmath}_{n+1})
\big]
\,,\nonumber
\end{eqnarray}
with the energy density per rung $E_{0}=-2/3$. (We recall that
symmetric ordering of spin-1 operators is implicitly assumed). For
this model two more eigenvalues of the local Hamiltonian are zero:
\[
\lambda_{0}=\lambda_{1}=0\,,
\]
which may in principle indicate higher degeneracy of the ground state
(there always exists another dimerized singlet state which can be
obtained by the translation in one rung, and $\lambda_{1}=0$ may
mean degeneracy with some ``partially ferromagnetic'' state with
the total spin $J=1$ of each plaquette).

(b) Setting $u=-{1\over2}\sqrt{2}$, $y=0$, and choosing $x={4\over9}$
to fix the scale, one obtains a model of $S=1$ and
$S={1\over2}$ chains coupled with {\em purely biquadratic\/}
interaction:
\begin{eqnarray} 
\label{biquad1} 
\widehat{H}&=&\sum_{n}
\mbox{\boldmath$S$\unboldmath}_{n} 
\mbox{\boldmath$S$\unboldmath}_{n+1}
+
{8\over3}
\mbox{\boldmath$\tau$\unboldmath}_{n}
\mbox{\boldmath$\tau$\unboldmath}_{n+1}\\
&+&(\mbox{\boldmath$S$\unboldmath}_{n}
\mbox{\boldmath$S$\unboldmath}_{n+1}) 
\big[
\mbox{\boldmath$\tau$\unboldmath}_{n}
\mbox{\boldmath$\tau$\unboldmath}_{n+1}
+(\mbox{\boldmath$S$\unboldmath}_{n}
+\mbox{\boldmath$S$\unboldmath}_{n+1})\cdot
(\mbox{\boldmath$\tau$\unboldmath}_{n}
+\mbox{\boldmath$\tau$\unboldmath}_{n+1})
\big]\,,
\nonumber
\end{eqnarray} 
with the energy per rung $E_{0}=-2$. For this model also one of the
local Hamiltonian eigenvalues vanishes,
\[
\lambda_{2}^{(1,1)}=0\,.
\]

Another model of this type can be obtained by setting
$u=-{1\over4}\sqrt{2}$, $y=0$; after choosing $x={8\over27}$ to fix a
proper scale the Hamiltonian takes the form
\begin{eqnarray} 
\label{biquad2} 
\widehat{H}&=& \sum_{n} 
\mbox{\boldmath$S$\unboldmath}_{n} 
\mbox{\boldmath$S$\unboldmath}_{n+1}
+
2\mbox{\boldmath$\tau$\unboldmath}_{n}
\mbox{\boldmath$\tau$\unboldmath}_{n+1}
+\left(
\mbox{\boldmath$S$\unboldmath}_{n}\mbox{\boldmath$S$\unboldmath}_{n+1}
\right)^{2}\nonumber\\
&+& {1\over4} 
(\mbox{\boldmath$S$\unboldmath}_{n}
\mbox{\boldmath$S$\unboldmath}_{n+1}) 
\big[
(\mbox{\boldmath$S$\unboldmath}_{n}
+\mbox{\boldmath$S$\unboldmath}_{n+1})
\cdot
(\mbox{\boldmath$\tau$\unboldmath}_{n}+
\mbox{\boldmath$\tau$\unboldmath}_{n+1})
\big] \\
&-&\big[
(\mbox{\boldmath$S$\unboldmath}_{n}+\mbox{\boldmath$S$\unboldmath}_{n+1})
\cdot\mbox{\boldmath$\tau$\unboldmath}_{n}
\big]
\big[
(\mbox{\boldmath$S$\unboldmath}_{n}+\mbox{\boldmath$S$\unboldmath}_{n+1})
\cdot\mbox{\boldmath$\tau$\unboldmath}_{n+1}
\big]\,.\nonumber
\end{eqnarray} 
The energy per rung is $E_{0}=-1$, and the eigenvalue $\lambda_{0}=0$.

(c) If $x=0$, then one can somewhat surprisingly observe that the
solution (\ref{sol2x3gen}) does not depend on the parameter $u=a/b$
entering the MP wavefunction (\ref{ansatz}).  Three of the local
Hamiltonian eigenvalues are now zero,
\[
\lambda_{0}=0,\quad \lambda_{2}^{(1,1)}=0,\quad \lambda_{3}=0\,,
\]
and one has to put $y<0$ for the remaining ones to be
positive. Setting the energy scale by fixing $y=-{1\over8}$, one
obtains the model
\begin{eqnarray} 
\label{cont1} 
\widehat{H}&=& \sum_{n}
\mbox{\boldmath$\tau$\unboldmath}_{n}
\mbox{\boldmath$\tau$\unboldmath}_{n+1}
+{1\over8} 
\mbox{\boldmath$S$\unboldmath}_{n}
\mbox{\boldmath$S$\unboldmath}_{n+1}\\
&-&{1\over8} 
\mbox{\boldmath$S$\unboldmath}_{n}
(2\mbox{\boldmath$\tau$\unboldmath}_{n}
+\mbox{\boldmath$\tau$\unboldmath}_{n-1}
+\mbox{\boldmath$\tau$\unboldmath}_{n+1}) 
+ {1\over2}
(\mbox{\boldmath$S$\unboldmath}_{n}
\mbox{\boldmath$S$\unboldmath}_{n+1})
(\mbox{\boldmath$\tau$\unboldmath}_{n}
\mbox{\boldmath$\tau$\unboldmath}_{n+1})\nonumber\\
&-&{1\over2}
\big[
(\mbox{\boldmath$S$\unboldmath}_{n}+\mbox{\boldmath$S$\unboldmath}_{n+1})
\cdot\mbox{\boldmath$\tau$\unboldmath}_{n}
\big]
\big[
(\mbox{\boldmath$S$\unboldmath}_{n}+\mbox{\boldmath$S$\unboldmath}_{n+1})
\cdot\mbox{\boldmath$\tau$\unboldmath}_{n+1}
\big]\,,\nonumber
\end{eqnarray}
with the energy density $E_{0}=-{1\over4}$ per rung. The fact that
$\lambda_{3}=0$ means that the singlet ground state of the model is
degenerate with the fully polarized ferromagnetic state: the
ferromagnetic state is the eigenstate of the Hamiltonian, and it is
a straightforward exercise to check that it has the same energy.
However, the ground state degeneracy of the model (\ref{cont1}) is
much higher: {\em any\/} wave function $|\Psi(u)\rangle$ of the form
(\ref{mp}), (\ref{ansatz}) with {\em arbitrary\/} parameter $u$ is the
ground state. One can easily calculate the overlap between two such
wave functions having different values of $u$:
\begin{equation} 
\label{overlap} 
\langle \Psi(u_{1})| \Psi(u_{2})\rangle=q^{N},\;\;
q={(1+u_{1}u_{2})^{2}\over (1+u_{1}^{2})(1+u_{2}^{2})}\leq1\,,
\end{equation} 
i.e., the two g.s.\ wave functions with different values of $u$ are
{\em asymptotically orthogonal\/} in thermodynamic limit $N\to\infty$
with the overlap vanishing exponentially with the increase of $N$.
This means that the dimension of the basis of this subspace
$\{|\Psi(u)\rangle\}$, i.e., the number of mutually orthogonal ground
states, is exponentially large in thermodynamic limit. This is an
example of {\em multicritical\/} model. 
Unfortunately, within the present approach one cannot make any statement
about presence of the gap.
It is worthwhile to mention that infinitely degenerate ground state in
a mixed spin-$1$/spin-${1\over2}$ system
was observed in certain limiting case of de Vega-Woynarovich model
(see the discussion of $\bar{c}=0$ case in Ref.\
\onlinecite{DeVega+94}) when the velocity of one of the two spinon
branches with linear dispersion becomes zero; however, since a pair of
spinons can be combined either in a singlet or in a triplet with the
same energy, the set of degenerate ground states in that case should
contain not only singlets, but states of higher spin as well.

\subsubsection{Case $b=0$ (no spin-${3\over2}$ states on the rung)}

For $2\times3$ MP ansatz it is useful to consider separately the case
$b=0$, because it turns out to lead to a new type of solution. One can
see that at $b=0$ the spin-${3\over2}$ states of the ladder rungs are
excluded from the wavefunction (\ref{ansatz}), and only the following
two multiplets are present in the matrix products
$g^{L}_{1}g^{R}_{2}$, $g^{R}_{1}g^{L}_{2}$:
\begin{equation} 
\label{b0lgs} 
|\Psi_{00}^{g}\rangle=|\psi^{11}_{00}\rangle\,,\quad
|\Psi_{1M}^{g}\rangle=|\psi^{11}_{1M}\rangle\,.
\end{equation}
It turns out that in this case it is possible to obtain a nontrivial
solution of the system (\ref{zeroeqs}) with {\em only bilinear\/}
coupling, which correspond to the following model:
\begin{eqnarray} 
\label{bilin} 
\widehat{H}&=&\sum_{n}
\gamma 
\mbox{\boldmath$S$\unboldmath}_{n}\mbox{\boldmath$\tau$\unboldmath}_{n}
-\mbox{\boldmath$\tau$\unboldmath}_{n}
\mbox{\boldmath$\tau$\unboldmath}_{n+1}
-\mbox{\boldmath$S$\unboldmath}_{n}
\mbox{\boldmath$S$\unboldmath}_{n+1}\nonumber\\
&-&\mbox{\boldmath$S$\unboldmath}_{n}
(\mbox{\boldmath$\tau$\unboldmath}_{n-1}
+\mbox{\boldmath$\tau$\unboldmath}_{n+1})\,.
\end{eqnarray}
Here $\gamma\geq {4\over3}$ is an arbitrary parameter, and the energy
density is $E_{0}=-(\gamma+{1\over4})$ per rung. At
$\gamma={4\over3}$ the eigenvalue $\lambda_{3}$ vanishes, indicating
the first-order transition into fully polarized ferromagnetic state
[cf.\ a similar transition for $S={1\over2}$ ladder in Ref.\
\onlinecite{KM97}]. 
The matrix product $g^{L}_{1}g^{R}_{2}$ has the simple form
\begin{equation} 
\label{vizb0}
\left[
\begin{array}{cc}
2|\Uparrow\Downarrow\rangle-|\Downarrow\Uparrow\rangle &
|\Downarrow\Downarrow\rangle \\
-|\Uparrow\Uparrow\rangle & -2|\Downarrow\Uparrow\rangle +
|\Uparrow\Downarrow\rangle 
\end{array}
\right]\,,
\end{equation}
which allows one to ``visualize'' the structure of the ground state as
interpolating between two ``U-shape'' VBS states shown in Fig.\
\ref{fig:VBS}c,d [here $\Uparrow$, $\Downarrow$ are the rung states
with total spin ${1\over2}$, see Eqs.\ (\ref{rung-states})].

\subsection{${\mathbf 2\times4}$ MP ansatz}
\label{subsec:2x4}

For the $2\times4$ MP ansatz, at a general choice of parameters $a$,
$b$, $w$ entering the wave function (\ref{ansatz}), the following
multiplets are contained in the two matrix products $g^{L}g^{R}$ and
$g^{R}g^{L}$:
\begin{eqnarray} 
\label{lgs2x4}
&& \psi^{11}_{00},\;\; \psi^{33}_{00},\nonumber\\
&& \psi^{11}_{1M},\;\; \psi^{33}_{1M},\;
\psi^{13}_{1M},\;\; \psi^{31}_{1M},\nonumber\\
&&(1/\sqrt{2})(\psi^{31}_{2M} - \psi^{13}_{2M}),\;\;
\psi^{33}_{2M}\,. 
\end{eqnarray}
We require them to be annihilated by the local Hamiltonian
$\widehat{h}$, and the remaining multiplets
\begin{equation} 
\label{les2x4} 
(1/\sqrt{2})(\psi^{31}_{2M}-\psi^{13}_{2M}),\;\;
\psi^{33}_{3M}
\end{equation}
to be the eigenstates of $\widehat{h}$ with positive energy. Eqs.\
(\ref{zeroeqs}) give a system of eighteen linear equations for twenty
parameters of the Hamiltonian (\ref{ham}), and its general solution
contains two free parameters $x$, $y$ (one of them is again irrelevant
since it just sets the energy scale). The multiplets
(\ref{lgs2x4}) do not contain the wave function parameters $a$, $b$,
$w$, and thus the solution also does not depend on them, yielding a
{\em one-parametric family of multicritical models\/}
with infinitely degenerate ground state similar to one discussed in
the previous subsection. The solution is given by
\begin{eqnarray} 
\label{sol2x4gen} 
&& J_{r}=J_{r}'=2x, \quad 
J_{d}=J_{d}'=K_{1}=K_{1}'=K_{2}=K_{2}'=x\,,\nonumber\\
&& J_{S}={3\over2}y,\quad J_{\tau}={4\over7}(5y-4x)\,,\nonumber\\
&& K_{S}={1\over2}y,\quad K_{S\tau}={2\over7}(13y-9x)\,,\\
&& K_{rr}=K_{dd}=3x-2y,\quad
K_{rd}=K_{rd}'={4\over7}(3x-2y)\,,\nonumber\\
&& U_{1}={2\over7}(5y-4x),\quad U_{2}=U_{3}={4\over7}(3x-2y)\,,\nonumber
\end{eqnarray}
and the conditions (\ref{lambdaeqs}) take the form
\begin{equation} 
\label{lambda2x4gen} 
\lambda_{2}=8(y-x)\geq 0,\qquad
\lambda_{3}={20\over7}(2x+y) \geq 0\,.
\end{equation}
This class of solutions is more cumbersome than in case of the
$2\times3$ MP ansatz; for instance, one may observe that six-spin
interactions $U_{1,2,3}$ in (\ref{sol2x4gen}) should be always
nonzero.  The simplest model within this class is achieved by setting
$y={3\over2}x$, its Hamiltonian after fixing the proper energy scale
is
\begin{eqnarray} 
\label{cont2} 
\widehat{H}&=& \sum_{n} 
\mbox{\boldmath$S$\unboldmath}_{n} 
\mbox{\boldmath$S$\unboldmath}_{n+1}
+{1\over3} \left(\mbox{\boldmath$S$\unboldmath}_{n} 
\mbox{\boldmath$S$\unboldmath}_{n+1}\right)^{2}
\big[
1+{4\over3}\mbox{\boldmath$\tau$\unboldmath}_{n}
\mbox{\boldmath$\tau$\unboldmath}_{n+1}
\big]\nonumber\\
&+&{8\over9}
\mbox{\boldmath$\tau$\unboldmath}_{n}
\mbox{\boldmath$\tau$\unboldmath}_{n+1}
+{4\over9}
\mbox{\boldmath$S$\unboldmath}_{n}
\big(2\mbox{\boldmath$\tau$\unboldmath}_{n}
+\mbox{\boldmath$\tau$\unboldmath}_{n-1}
+\mbox{\boldmath$\tau$\unboldmath}_{n+1}\big)
\\ 
&+&{4\over9}
(\mbox{\boldmath$S$\unboldmath}_{n}
\mbox{\boldmath$S$\unboldmath}_{n+1}) 
\big[
(\mbox{\boldmath$S$\unboldmath}_{n}
+\mbox{\boldmath$S$\unboldmath}_{n+1})\cdot
(\mbox{\boldmath$\tau$\unboldmath}_{n}
+\mbox{\boldmath$\tau$\unboldmath}_{n+1})
\big]\,,
\nonumber
\end{eqnarray}
with the energy per rung $E_{0}=-{2\over3}$.  It should be remarked
that, in contrast to the model (\ref{cont1}), fully polarized
ferromagnetic state  generally (except for the case $x=-{1\over2}y$)
is not degenerate with the ground state of multicritical models defined
by (\ref{sol2x4gen}).

\section{Summary}
\label{sec:summary}

We have studied the isotropic ladder composed of $S=1$ and
$S={1\over2}$ chains, with general type exchange interaction between
spins on neighboring rungs. The technique of matrix product states is
applied to construct a family of Hamiltonians with exact ground
states.  Among the members of this family, we have found a couple of
relatively simple models with nontrivial ground states, including one
model with only bilinear exchange and two models with $S=1$ and
$S={1\over2}$ chains being coupled by purely biquadratic exchange. We
also present a family of multicritical models whose ground state is
infinitely degenerate: in the thermodynamic limit $N\to\infty$ the
number of degenerate ground states grows with $N$ exponentially.

\acknowledgements 

We are grateful to H.-U. Everts and C. Waldtmann for
discussion of the results.  A.K. gratefully acknowledges the
hospitality of Hannover Institute for Theoretical Physics during his
stay there.  This work was supported by the German Ministry for
Research and Technology (BMBF) under the contract 03MI4HAN8 and by the
Ukrainian Ministry of Science (grant 2.4/27).

\appendix

\section{Technical details}
\label{app:tech}

Here is the explicit form of the matrices we used in (\ref{ansatz}) for
the $2\times3$ MP ansatz:
\begin{eqnarray} 
\label{tjm2x3} 
&& T_{L}^{ {1\over2},{1\over2} }=
\left[
\begin{array}{lcr} 
0 & \sqrt{{1\over3}} & 0  \\
0 & 0          & \sqrt{{2\over3}} 
\end{array}
\right]
,\;
T_{L}^{{1\over2},-{1\over2}}=
\left[
\begin{array}{lcr}
-\sqrt{{2\over3}} & 0           & 0  \\
0          & -\sqrt{{1\over3}} & 0
\end{array}
\right]
,\nonumber\\
&&T_{L}^{{3\over2},{3\over2}}=
\left[
\begin{array}{lcr}
0 & 0 & 1  \\
0 & 0 & 0
\end{array}
\right]
,\quad
T_{L}^{{3\over2},-{3\over2}}=
\left[
\begin{array}{lcr}
0 & 0 & 0  \\
1 & 0 & 0
\end{array}
\right]
,\\
&&T_{L}^{{3\over2},{1\over2}}=
\left[
\begin{array}{lcr}
0 & -\sqrt{{2\over3}} & 0  \\
0 & 0          & \sqrt{{1\over3}}
\end{array}
\right]
,\;
T_{L}^{{3\over2},-{1\over2}}=
\left[
\begin{array}{lcr}
\sqrt{{1\over3}} & 0           & 0  \\
0          & -\sqrt{{2\over3}} & 0
\end{array}
\right].\nonumber
\end{eqnarray}
The matrices $T_{R}$ can be easily obtained from those matrices using
the definition (\ref{TR}).

And in case of the $2\times4$ ansatz the matrices were chosen as
\begin{eqnarray} 
\label{tjm2x4}
&&T_{L}^{{1\over2},{1\over2},0}= 
\left[
\begin{array}{lccr}  
0 & 1/\sqrt{2} & -1/\sqrt{2} & 0 \\
0 & 0 & 0 & 0
\end{array}
\right]\,,
\nonumber\\
&&T_{L}^{{1\over2},-{1\over2},0}= 
\left[
\begin{array}{lccr}  
0 & 0 & 0 & 0\\
0 & 1/\sqrt{2} & -1/\sqrt{2} & 0 
\end{array}
\right]\,,
\nonumber\\
&&T_{L}^{{1\over2},{1\over2},1}= 
\left[
\begin{array}{lccr}  
0 & 1/\sqrt{6} & 1/\sqrt{6} & 0 \\
0 & 0 & 0 & \sqrt{2/3}
\end{array}
\right]\,,
\nonumber\\
&&T_{L}^{{1\over2},-{1\over2},1}= 
\left[
\begin{array}{lccr}  
-\sqrt{2/3} & 0 & 0 & 0 \\
0 & -1/\sqrt{6} & -1/\sqrt{6} & 0 
\end{array}
\right]\,,\\
&&T_{L}^{{3\over2},{3\over2}}= 
\left[
\begin{array}{lccr}  
0 & 0 & 0 & 1 \\
0 & 0 & 0 & 0
\end{array}
\right]\,,\quad
T_{L}^{{3\over2},-{3\over2}}= 
\left[
\begin{array}{lccr}  
0 & 0 & 0 & 0 \\
1 & 0 & 0 & 0
\end{array}
\right]\,,
\nonumber\\
&&T_{L}^{{3\over2},{1\over2}}= 
\left[
\begin{array}{lccr}  
0 & -1/\sqrt{3} & -1/\sqrt{3} & 0 \\
0 & 0 & 0 & 1/\sqrt{3}
\end{array}
\right]\,,
\nonumber\\
&&T_{L}^{{3\over2},{1\over2}}= 
\left[
\begin{array}{lccr}  
1/\sqrt{3} & 0 & 0 & 0 \\
0 & -1/\sqrt{3} & -1/\sqrt{3} & 0 
\end{array}
\right]\,.\nonumber
\end{eqnarray}

The complete set of spin states of a two-rung plaquette is, in the
notation used in
Eqs. (\ref{rung-states}),
\begin{eqnarray}
\label{plaquette-states}
&&\psi_{33}^{33}=|33\rangle,\quad
  \psi_{3,-3}^{33}=|\bar{3}\bar{3}\rangle,\nonumber\\
&&\psi^{33}_{32}={1\over\sqrt{2}}(|31\rangle
        +|13\rangle),\quad
   \psi^{33}_{3,-2}={1\over\sqrt{2}}(|\bar{3}\bar{1}\rangle
        +|\bar{1}\bar{3}\rangle),\nonumber\\
&&\psi^{33}_{3,1}=(1/\sqrt{5})(\sqrt{3}|11\rangle
        +|3\bar{1}\rangle
        +|\bar{1}3\rangle),\nonumber\\
&&\psi^{33}_{3,-1}=(1/\sqrt{5})(\sqrt{3}|\bar{1}\bar{1}\rangle
        +|\bar{3}1\rangle
        +|1\bar{3}\rangle),\nonumber\\
&&\psi^{33}_{30}=(1/2\sqrt{5})\{|3\bar{3}\rangle
        +|\bar{3}3\rangle
        +3(|1\bar{1}\rangle+|\bar{1}1\rangle)\},
        \nonumber\\
&&\psi^{33}_{22}={1\over\sqrt{2}}(|31\rangle
                   -|13\rangle),\quad
  \psi^{33}_{2,-2}=-{1\over\sqrt{2}}(|\bar{3}\bar{1}\rangle
                   -|\bar{1}\bar{3}\rangle),\nonumber\\
&&\psi^{33}_{21}={1\over\sqrt{2}}(|3\bar{1}\rangle
                   -|\bar{1}3\rangle),\quad
  \psi^{33}_{2,-1}=-{1\over\sqrt{2}}(|\bar{3}1\rangle
                   -|1\bar{3}\rangle),\nonumber\\
&&\psi^{33}_{20}=(1/2)(|1\bar{1}\rangle
         -|\bar{1}1\rangle +|3\bar{3}\rangle
         -|\bar{3}3\rangle),\nonumber\\
&&\psi^{33}_{11}=\sqrt{{3/10}}(|3\bar{1}\rangle
             +|\bar{1}3\rangle
             -(2/\sqrt{3})|11\rangle),\nonumber\\
&&\psi^{33}_{1,-1}=\sqrt{{3/10}}(|\bar{3}1\rangle
             +|1\bar{3}\rangle
             -(2/\sqrt{3})|\bar{1}\bar{1}\rangle),\nonumber\\
&&\psi^{33}_{10}=(1/2\sqrt{5})\{3(|3\bar{3}\rangle
             +|\bar{3}3\rangle)
             -|1\bar{1}\rangle
             -|\bar{1}1\rangle \},\\
&&\psi^{33}_{00}=(1/2)(|\bar{1}1\rangle
         -|1\bar{1}\rangle +|3\bar{3}\rangle
         -|\bar{3}3\rangle),\nonumber\\
&&\psi^{31}_{22}=|3\Uparrow\rangle,\quad
  \psi^{31}_{2,-2}=|\bar{3}\Downarrow\rangle,\nonumber\\
&&\psi^{31}_{21}={1\over2}|3\Downarrow\rangle
          +{\sqrt{3}\over2}|1\Uparrow\rangle,\quad
  \psi^{31}_{2,-1}={1\over2}|\bar{3}\Uparrow\rangle
          +{\sqrt{3}\over2}|\bar{1}\Downarrow\rangle,\nonumber\\
&&\psi^{31}_{20}={1\over\sqrt{2}}(|1\Downarrow\rangle
          +|\bar{1}\Uparrow\rangle),\quad
  \psi^{31}_{11}=-{1\over2}|1\Uparrow\rangle
          +{\sqrt{3}\over2}|3\Downarrow\rangle,\nonumber\\
&&\psi^{31}_{1,-1}={1\over2}|\bar{1}\Downarrow\rangle
          -{\sqrt{3}\over2}|\bar{3}\Uparrow\rangle,\quad
  \psi^{31}_{10}={1\over\sqrt{2}}(|1\Downarrow\rangle
          -|\bar{1}\Uparrow\rangle), \nonumber\\
&&\psi^{11}_{00}={1\over\sqrt2}
\big(|\Uparrow\Downarrow\rangle-|\Downarrow\Uparrow\rangle\big),
\quad \psi^{11}_{10}={1\over\sqrt2}
\big(|\Uparrow\Downarrow\rangle+
|\Downarrow\Uparrow\rangle\big),\nonumber\\
&&\psi^{11}_{11}=|\Uparrow\Uparrow\rangle,
\psi^{11}_{1,-1}=
|\Downarrow\Downarrow\rangle\,.\nonumber
\end{eqnarray}
Here the superscripts of $\psi$'s denote the total momentum $J$ of
left and right rung states of which they are composed (with the shortcut
convention that $3$ means $J={3\over2}$ and $1$ means ${1\over2}$).
The states $\psi^{13}_{jm}$ can be obtained from $\psi^{31}_{jm}$ by
interchanging the left and right rung states.

\section{General solution for the $\mathbf 2\times3$ MP ansatz}
\label{app:gen}

The general solution of the system (\ref{eqn2x3gen}) with four
additional conditions $\lambda_{2}^{(1,2)}=0$, $U_{1,2,3}=0$ has the
following form:
\begin{mathletters}
\label{sol2x3gen}
\FL
\begin{eqnarray}
\label{zero2x3gen}
&&J_{S} = 
 \big[  15u^{4}  
+ 37u^{2} +  {5\over2}  -\sqrt{2} u(2u^{4}-40u^{2}-3)
 \big] x - y\,,\nonumber\\
&&J_{\tau} =  (  112u^{4} + 176u^{3}
\sqrt {2} + 192u^{2} + 44\sqrt {2}u + 14   ) 
x - 8y\,,\nonumber\\
&&J_{r} =  2\sqrt {2}
 (  4u + \sqrt{2}  ) (u^{2} + 1)
 (  2u^{2} -1  )  x + 2y\,,\nonumber\\
&&J_{d} = y\,,\\
&&K_{S} =  - \frac {1}{2}\sqrt {2}
 (  2u + \sqrt {2}   )  (  {u
} + \sqrt {2}   ) u (   - 2u + 
\sqrt {2}   ) ^{2} x\,,\nonumber\\
&&K_{S\tau} = 4 \big[ 13u^{4} + 25u^{2} + 3/2\nonumber\\
&&\qquad\qquad\qquad\qquad
 -\sqrt{2}u(2u^{4}-24u^{2}-5) \big] x - 4y \,,\nonumber\\
&&K_{rr} =  - \sqrt {2} (   - 4u^{4} + 11
u^{3}\sqrt {2} + 21u^{2} + 7\sqrt {2}u + 5  
 ) \nonumber\\
&&\qquad\qquad\qquad\qquad\qquad\qquad
\times (  2u + \sqrt {2}   ) x + 4y \,,\nonumber\\
&&K_{dd} =  - \sqrt {2} (  4u^{4} + 9u^{3}\sqrt {2} 
+ 27u^{2} + 5\sqrt {2}u + 3  
 )  \nonumber\\
&&\qquad\qquad\qquad\qquad\qquad\qquad
\times(  2u + \sqrt {2}   ) x + 4y \,,\nonumber\\
&&K_{rd} =  - 4 (  5u^{2} + \sqrt {2}u
 + 1   )  (  u + \sqrt {2}   ) \nonumber\\
&&\qquad\qquad\qquad\qquad\qquad\qquad
\times (  2u + \sqrt {2}   ) x + 4y \,,\nonumber\\
&&K_{1} =  -  (   - 2u + \sqrt {2}  
 )  (  u + \sqrt {2}   )  ( 
 u^{2} + 3\sqrt {2}u + 1   ) x \,,\nonumber\\
&&K_{2} = \frac {1}{2}\sqrt {2}
 (   - 2u + \sqrt {2}   )  (  8
u^{3} + 5u^{2}\sqrt {2} + 2u - \sqrt {2}  
 ) u x \,,\nonumber\\
&&E_{0} =  -\big[  
44u^{4} + 68u^{2} + 9/2\nonumber\\
&&\qquad\qquad\qquad\qquad + \sqrt{2}u(8u^{4}+60u^{2}+13)
\big] x  
+ 2y \,;\nonumber
\end{eqnarray} 
with the conditions on the eigenvalues (\ref{lambdaeqs})
 being
\begin{eqnarray}
\label{lambda2x3gen}
&&\lambda_{0} =  - 6\sqrt {2} (4u + \sqrt {2} ) x \geq 0
\,,\nonumber\\
&&\lambda_{1} =  (  48u^{4} + 66u^{3}
\sqrt {2} + 78u^{2} + 18\sqrt {2}u + 6   ) 
x - 4y \geq0\,,\nonumber\\
&&\lambda_{2}^{(1,1)} = 3\sqrt {2} (  2u + 
\sqrt {2}   ) (u^{2} + 1) x \geq0\,,\\
&&\lambda_{2}^{2, 2} = 12 (  12u^{3} 
+ 9u^{2}\sqrt {2} + 4u + \sqrt {2}   ) \nonumber\\
&&\qquad\qquad\qquad\qquad\qquad\qquad
\times (  
u + \sqrt {2}   ) x - 12y \geq 0 \,,\nonumber\\
&&\lambda_{3} = 60u^{2} (  u + \sqrt {2}
   ) ^{2} x\geq0 \,.\nonumber
\end{eqnarray}
\end{mathletters}
Here $u=a/b$ is the free parameter entering the $2\times3$ MP ansatz
(\ref{ansatz}), and $x,y$ are additional free parameters arising from
the solution of the linear system (\ref{eqn2x3gen}). Taking into
account that one of the parameters just sets the energy scale and thus
is irrelevant, one gets from (\ref{sol2x3gen}) a two-parameter family
of Hamiltonians with exact ground states in a form of the $2\times3$
matrix product.  One can see that inequalities (\ref{lambda2x3gen})
can be satisfied provided that
\begin{equation} 
\label{ranges2x3}
 x\geq 0\,,\qquad -{\sqrt{2}\over4}\leq u\leq -{\sqrt{2}\over2}\,. 
\end{equation}

\end{multicols}

\newpage

\vskip 10mm
\end{document}